\documentclass{aa}  

\usepackage{graphicx}
\usepackage{txfonts}
\usepackage{hyperref}
\hypersetup{
    colorlinks   = true, %Colours links instead of ugly boxes
    urlcolor     = cyan, %Colour for external hyperlinks
    linkcolor    = blue, %Colour of internal links
    citecolor   = blue %Colour of citations
}
\usepackage{placeins}
\usepackage{xcolor}
\usepackage{tabularray} % for multi-columns tables

\graphicspath{plots}

\begin{document}

\title{
The most stringent upper limit from dynamical models on the mass of a central black hole in 47~Tucanae
}
\titlerunning{Upper limit on the IMBH mass in 47~Tucanae}

\author{
A. Della Croce \inst{1,2} \thanks{\email{alessandro.dellacroce@inaf.it}} \and
R. Pascale \inst{2} \and 
E. Giunchi \inst{3,4} \and
C. Nipoti \inst{1} \and
M. Cignoni \inst{5,6} \and
E. Dalessandro \inst{2}
}
\institute{
Dipartimento di Fisica e Astronomia "Augusto Righi", Universit\`a di Bologna, via Piero Gobetti 93/2, I-40129 Bologna, Italy
\and
INAF - Osservatorio di Astrofisica e Scienza dello Spazio di Bologna, via Piero Gobetti 93/3, I-40129 Bologna, Italy \and
INAF-Osservatorio astronomico di Padova, Vicolo Osservatorio 5, 35122 Padova, Italy \and 
Dipartimento di Fisica e Astronomia, Universit\`a di Padova, Vicolo Osservatorio 3, 35122 Padova, Italy \and 
Physics Departement, University of Pisa, Largo Bruno Pontecorvo, 3, I-56127 Pisa, Italy \and 
INFN, Largo B. Pontecorvo 3, 56127, Pisa, Italy
}
\authorrunning{Della Croce et al.}
\date{Received ...; accepted ...}
 
\abstract
{
Globular clusters (GCs) were proposed as promising sites for discovering intermediate-mass black holes (IMBHs), possibly providing crucial insights into the formation and evolution of these elusive objects. 
The Galactic GC 47~Tucanae (also known as NGC~104) has been suggested as a potential IMBH host, but, previous studies have yielded conflicting results.
We, therefore, present self-consistent dynamical models based on distribution functions (DFs) that depend on action integrals to assess the presence (or absence) of an IMBH in 47~Tucanae. 
Leveraging state-of-the-art Multi Unit Spectroscopic Explorer and Hubble Space Telescope data, we analyzed the three-dimensional (3D) kinematics of the cluster's central regions, fitting individual star velocities down to the sub-arcsec scale (approximately $10^{-2}$ pc).
According to our analysis, the inner kinematics of 47~Tucanae is incompatible with a central BH more massive than 578~M$_\odot$ (at $3\sigma$).
This is the most stringent upper limit on the mass of a putative IMBH in 47~Tucanae that has been put by any dynamical study.
}


\maketitle

\section{Introduction}
Intermediate-mass black holes (IMBHs) are classified as those black holes (BHs) with mass in the range $10^2-10^5$ M$_\odot$ \citep{Greene_etal2020}, i.e.\ between those of stellar-mass BHs and of supermassive BHs (SMBHs).
%($M_\bullet <10^2$ M$_\odot$)  , $M_\bullet>10^6$ M$_\odot$
The discovery of SMBHs at $z=7.5$, when the Universe was only $0.7$~Gyr old \citep{Banados_etal2018}, poses a challenge to theories of SMBH formation \citep{Volonteri_2010}, and, since IMBHs are thought to be possible seeds from which SMBHs grew at early times, finding evidence for IMBHs would provide insights into BH formation mechanisms.
However, so far we have no firm evidence of BHs in the range $10^2-10^5$ M$_\odot$ \citep[see e.g.][]{Brok_2015,Nguyen_etal2018,abbott_etal2020}.

GCs are good candidates to host IMBHs, because (i) theoretically they are expected to be promising sites for IMBH formation \citep{MillerHamilton_2002,Portegies-Zwart_etal2004} and (ii) empirically central IMBHs are predicted in GCs
as a natural extrapolation of the \citet{Magorrian_etal1998} relation 
between central BH mass and bulge mass observed in galaxies.
The presence of IMBHs in GCs has been investigated by means of different techniques, such as radio emission 
% \cn{ho cambiato gas dynamical studies in radio emission: e' OK?} 
\citep[e.g.][]{Strader_etal2012,Tremou_2018}, kinematic studies of the innermost stars \citep[e.g.][]{Gerssen_etal2002,vitral_etal2023}, and studies that constrain the gravitational field using the timing of radio pulsars \citep[e.g][]{Kiziltan_etal2017,Abbate_etal2018}.

The Galactic GC 47~Tucanae is arguably one of the best targets to look for an IMBH, mainly because of its high density and mass \citep{MillerHamilton_2002, Portegies-Zwart_etal2004, Giersz_etal2015}. It is also relatively nearby ($\simeq 4.5$ kpc), which allows for detailed studies of the central kinematics.
% \cn{completare... "of its high central density and mass (?)" Le motivazioni sono teoriche (e' denso quindi si puo' essere formato il black hole), pratiche (e' vicino, luminoso ?) o entrambe?}  
It was thus studied with different approaches to investigate the possible presence of an IMBH.
% If the IMBH were accreting mass, a fraction of the rest mass energy of the infalling gas would be converted into radiation. Such radiation is expected to be detectable at radio wavelengths \citep{deRijcke_etal2006}. 
Some studies carried out radio observations of the core of 47~Tucanae \citep[e.g.][]{deRijcke_etal2006, Tremou_2018}, finding no evidence of a significant emission.
\citet{Tremou_2018} put a 3$\sigma$ upper limit at $M_\bullet<1040$ M$_\odot$, while \citet{deRijcke_etal2006} found a broader limit $M_\bullet<670-2060$ M$_\odot$, depending on different assumptions on the gas density, gas temperature, and the fraction of rest-mass energy of the infalling matter converted into radiation. 
Comparing spin-down measurements for nineteen, millisecond pulsars (MSPs) identified in 47~Tucanae, \citet{Kiziltan_etal2017} found that an IMBH of mass $M_\bullet=2300^{+1500}_{-850}$ M$_\odot$ is required to reproduce the accelerations and the cumulative spatial distribution of MSPs.
\citet{HenaultBrunet_etal2020} found 
that a total mass of $430^{+386}_{-301}$ M$_\odot$ in stellar-mass BHs could explain the stellar kinematics and spatial distribution \citep[][without IMBH]{HenaultBrunet_etal2020}.
% by means of dynamical models that an IMBH is not needed to explain the properties of 47~Tucanae. 
% According to their result, a total mass of $430^{+386}_{-301}$ M$_\odot$ in stellar-mass BHs could explain the observations \citep{HenaultBrunet_etal2020}.
Exploiting a set of HST PM measurements of the central regions of 47~Tucanae, \citet{Mann_etal2020} found that
the stellar BH population cannot fully account for the observed velocity dispersion, even if a BH and neutron star retention fraction 
% (i.e. the fraction of objects that were not expelled out of the cluster due to natal kicks and/or dynamical interactions) 
of
the 100\% is assumed. They concluded that an additional massive component with a mass $M_\bullet=808-4710$ M$_\odot$ (depending on the retention fraction) is favored. 
% The result of \citet{Kiziltan_etal2017} is compatible with a BH retention fraction of 25\% in the model of \citet{Mann_etal2020}.
%%%%%%

The tension among some of the aforementioned results suggests that the question of the presence of an IMBH in 47~Tucanae requires further investigation. 
In this work, we address the problem by means of a stellar dynamical approach that combines state-of-the-art structural and kinematic data of 47 Tucanae and flexible self-consistent models of stellar systems with 
the optional presence of a central BH.

%that depend on the action integrals \citep[see e.g.][]{Pascale_etal2019}, and employing a star-by-star approach in the model to data comparison.
% This approach allows us to fully exploit the model based on DFs, which predicts the full shape of the velocity distribution, without losing the information encoded in the data.

\section{Dynamical models}
\label{sec:models}
% GCs are collisional systems over time scales comparable to their ages. Indeed, typical two-body relaxation time \citep[$t_{\rm rh}$][]{Spitzer_1987} varies between $5\times10^7$ yr and $4\times10^{10}$ yr \citep{Harris_2010}.
% However, when compared to dynamical time, $t_{\rm rh}$ is orders of magnitude larger.
% We can therefore describe states of equilibrium of the system through the collisionless Boltzmann equation \citep[CBE,][]{Hamilton_etal2018}. Furthermore, according to \citet{Jeans_1915} theorem, any function of the integrals of motion solves the steady-state CBE. 
In this work, we used dynamical models based on DFs that depend on the action integrals $\mathbf{J}$ \citep[see e.g.][]{arnold1989mathematical, BT2008}. 
Describing a stellar system as an ensemble of orbits, we can represent each orbit through its actions. 
Orbits with small $|\mathbf{J}|$ populate the internal regions of the clusters, whereas large $|\mathbf{J}|$ values describe orbits in the external regions \citep{BT2008}.
% Actions are a set of integrals of motion which, complemented by their conjugate variables, the angles, form a canonical set of the phase space. They are defined as
% \begin{equation}
%     J_i\equiv\frac{1}{2\pi}\oint_{\gamma_i}\,{\rm\mathbf{p}}\cdot d{\rm \mathbf{q}}\,,
% \end{equation}
% where $({\rm\mathbf{q}},{\rm\mathbf{p}})$ is a set of canonical coordinates and $\gamma_i$ is a closed path along which the corresponding angle performs a full oscillation of $2\pi$.
This approach has a few important advantages:
i) the model is physical since the DF is always non-negative by construction; ii) the velocity anisotropy, as well as any physical property of the system, are self-consistently computed directly from the DF (see Section~\ref{sec:observables_from_DF});
% In addition, DFs that depend on the action integrals could in principle be extended to rotating and flattened systems \citep{binney_2014}. 
iii) 
the extension to multi-component systems, e.g. galaxies with a stellar and dark matter component \citep{Piffl_etal2015, BinneyPiffl2015, Pascale_etal2018} or GCs with a central BH \citep{Pascale_etal2019}, is straightforward.

\subsection{Model for the stellar component}
We consider models where the stellar component of 47~Tucanae is described by the DF
\begin{align}
    f_\star(\mathbf{J})  = & f_0\,M_\star\,\left[ 1+\left(\frac{J_0}{h(\mathbf{J})} \right)^{\zeta}\right]^{\Gamma/\zeta} \times \left[ 1+\left(\frac{g(\mathbf{J})}{J_0} \right)^{\zeta}\right]^{-(B-\Gamma)/\zeta} \times \nonumber\\ 
    &\text{exp}\,{\left[-\left(\frac{g(\mathbf{J})}{J_\text{cut}} \right)^{\alpha} \right]}\,,
    \label{eq:DF}
\end{align}
which produces models whose spatial distributions closely follow a double-power law model (\citealt{vasiliev_2019}, but see also \citealt{evans2014, BinneyPiffl2015, Pascale_etal2018, Pascale_etal2019}) with an exponential cutoff in the system outskirts.
Here $f_0$ is such that the DF is normalized to the total stellar mass $M_\star=(2\pi)^3\,\int f_\star(\mathbf{J}) d^3\mathbf{J}$.

The dimensionless free parameters $\Gamma$ and $B$ mainly determine the inner ($|\mathbf{J}|\lesssim J_0$) and outer ($|\mathbf{J}|\gtrsim J_0$) slopes in the action space, with $J_0$ being the typical action at which this transition takes place.
In the case of the double power-law model, $\Gamma$ and $B$ can be converted in the slopes of the three-dimensional density profile \citep{Posti_etal2015}.
The transition regime ($|\mathbf{J}|\sim J_0$) is mainly regulated by $\zeta$. Finally, the parameter $\alpha$ controls the sharpness of the exponential truncation for $|\mathbf{J}|\gtrsim J_{\rm cut}$, with $J_{\rm cut}$ ($>J_0$) being the typical action value above which the exponential cutoff dominates the stellar distribution.
The functions $h(\mathbf{J})$ and $g(\mathbf{J})$ are linear combinations of the actions defined as
\begin{align}
    h(\mathbf{J}) &=(3-2h_z) J_r + h_z(J_z+|J_\phi|)\equiv(3-2h_z) J_r + h_z |\mathbf{L}|\,, \nonumber\\
    g(\mathbf{J})&=(3-2g_z) J_r + g_z(J_z+|J_\phi|)\equiv(3-2g_z) J_r + g_z |\mathbf{L}|\,,
    \label{eq:hJ_gJ_functions}
\end{align}
where $|\mathbf{L}|$ is the total angular momentum. These functions depend only on the two free parameters $h_z$ and $g_z$, which mainly control the inner and outer anisotropy of the system respectively \citep[see section~4.1 in][]{vasiliev_2019}.
% As $h(\mathbf{J})$ and $g(\mathbf{J})$ should be positive for every choice of actions to ensure that the DF could be normalized, $h_z$ and $g_z$ can vary between $0$ and $3/2$.  
% The parameters $h_z$ and $g_z$ mainly control the anisotropy of the system. 
% Increasing either of the two disfavors orbits with larger angular momentum, since the DF decreases quicker in $|\mathbf{L}|$ than in $J_r$, thereby favoring the development of radial anisotropy and vice-versa.
\subsection{The gravitational potential}
The total gravitational potential of the model cluster is the sum of the BH potential $\Phi_\bullet$ and the stellar potential $\Phi_\star$. The  BH potential is
\begin{equation}
    \Phi_\bullet(r)= - \frac{G\,M_\bullet}{r}\,,
\end{equation}
where $M_\bullet$ is the BH mass and $r$ is the radial spherical coordinate.
The stellar potential is determined by numerically solving (in an iterative fashion) the Poisson equation. At each iteration $i$, the stellar potential is updated according to
\begin{align}
    \nabla^2 \Phi_{\star,\,i+1} &= 4\pi G \,\rho_{\star,\,i} \nonumber\\
    &= 4\pi G \int d^3\boldsymbol{v}\, f_\star(\boldsymbol{J}[\boldsymbol{x},\boldsymbol{v}|\Phi_{\star,\,i} + \Phi_\bullet])\,,
\end{align}
where $\rho_\star = \int d^3\boldsymbol{v}\,f_\star(\boldsymbol{J})$ is the 3D stellar density, and we made explicit the dependence of the conversion between actions and Cartesian phase-space coordinates $(\boldsymbol{x},\boldsymbol{v})$ on the total potential $\Phi_{\star,i} + \Phi_\bullet$.
This shows that, given a DF, the stellar density distribution, as well as any physical property derived for the visible component (see Section~\ref{sec:observables_from_DF}), depends on the combination of stellar and BH potential. 
% This shows that both the spatial and velocity distributions of the stellar component are subject to the combination of stellar and BH potential.
As an initial guess on $\Phi_\star$ we adopted the isochrone potential \citep{BT2008}, but we note that the final stellar potential does not depend on this specific choice \citep[see][]{vasiliev_2019}.

Since the BH potential is spherically symmetric, and the $h(\mathbf{J})$ and $g(\mathbf{J})$ functions are defined such that the DF depends only on the radial action and the angular momentum modulus, the overall system is also spherical.
Any integral of the DF that involves conversion between actions and Cartesian phase-space coordinates is performed with the \texttt{AGAMA}\footnote{Action-based GAlaxy Modelling Architecture.} library \citep{vasiliev_2019}.

\subsection{Observable properties from a DF} \label{sec:observables_from_DF}
Given a DF, we can calculate the observable properties of the model by suitable integrations of the DF allowing us to compare theoretical models against the observations. 
Throughout this work, the DF is normalized to the total system mass (see eq.~\ref{eq:DF}).
The mass surface density distribution is then obtained 
through
\begin{equation}
    \Sigma_\star(R) = \int_{-\infty}^{+\infty} dz\,\rho_\star(r) \,,
    \label{eq:proj_density}
\end{equation}
where 
% $\rho_\star(r) \equiv \int d^3\mathbf{v}\,f_\star(\mathbf{J})$ is the 3D density, 
$z$ is the LOS direction, and $R^2=r^2-z^2$ is the distance from the GC center on the plane of the sky. 
The stellar number density, $n_\star$, is then simply defined by the relation $\Sigma_\star \equiv m\,n_\star$, where $m$ is a nuisance parameter of the model with the dimension of a mass.

Furthermore, we can calculate projected velocity distributions 
as
\begin{equation}
    \mathcal{V}_{\star {\rm 3D}}(\boldsymbol{v}_{\rm 3D}|R) \equiv 
    \frac{ \int dz\,f_\star(\mathbf{J})}{\Sigma_\star(R)}\,,
    \label{eq:3Dprojected_veldistr}
\end{equation}
where $\boldsymbol{v}_{\rm 3D} = \{ v_{\rm R}, v_{\rm T}, v_{\rm LOS}\}$ is the vector of 3D projected velocities (i.e. $v_{\rm R}$ and $v_{\rm T}$ on the plane of the sky, while $v_{\rm LOS}$ is the LOS component).
Since for the majority of the stars, the 3D velocity is not available, it is useful to define the marginalized velocity distributions.
In particular, the LOS velocity distribution
\begin{equation}
    \mathcal{V}_{\star\,{\rm LOS}}(v_{\rm LOS}|R) \equiv \frac{\int dz\, dv_{\rm R}\,dv_{\rm T}\,f_\star(\mathbf{J})}{\Sigma_\star(R)}\,,
    \label{eq:LOSVD}
\end{equation}
and the distribution in the plane-of-the-sky velocity components
\begin{equation}
    \mathcal{V}_{\star\,{\rm PM}}(v_{\rm R},\,v_{\rm T}|R) \equiv \frac{\int dz\, dv_{\rm LOS}\,f_\star(\mathbf{J})}{\Sigma_\star(R)}\,,
    \label{eq:PMVD}
\end{equation}
such that they are normalized to unity in the velocity space.
Finally, the velocity dispersion profile of the $i$-th velocity component (with $i=$"R", "T" or "LOS") is computed as
\begin{equation}
    \sigma^2_{\star\,i}(R) \equiv \frac{\int dz\,d^3\mathbf{v}\,v^2_i\,f_\star(\mathbf{J})}{\Sigma_\star(R)} \,.
    \label{eq:velocity_dispersion}
\end{equation}
Eq.~\ref{eq:proj_density} to \ref{eq:velocity_dispersion}
allow us to test theoretical predictions against the data (see Section~\ref{sec:data_analysis}).

\section{Data Analysis}
\label{sec:data_analysis}

The family of dynamical models presented in Section~\ref{sec:models} has eleven free parameters.
Namely, the total stellar and BH masses $M_\star$, and $M_\bullet$, the scale actions $J_0$, and $J_{\rm cut}$, the dimensionless free parameters $\zeta$, $\Gamma$, $B$, $g_{\rm z}$, $h_{\rm z}$, and $\alpha$, and the nuisance parameter $m$.
We explored this parameter space by comparing the models with a set of observables in a fully Bayesian framework. 
Details on the likelihood and the Markov Chain Monte Carlo used to explore the model posterior and calculate uncertainties on the free parameters (and on any derived quantity) are given in Appendix~\ref{sec:posterior_definition}.

As a kinematic dataset, we used a combination
% The 47~Tucanae kinematic dataset we used is a combination of the kinematic samples 
of individual LOS velocities from \citet[][obtained using the Multi Unit Spectroscopic Explorer, MUSE, spectrograph]{kamann+18} and PMs from \citet[][derived from multi-epoch observations with HST]{libralato+22}. 
% In addition, we used the density profile obtained by \citet{deBoer_etal2019}.

% We adopted the kinematic sample of individual LOS velocities from \citet{kamann+18} obtained using the MUSE spectrograph.
\citet{kamann+18} obtained individual LOS velocities using the MUSE spectrograph.
This sample represents the largest compilation of LOS velocities covering the central regions of 47~Tucanae (up to 100" from the GC center), with a typical velocity accuracy of $1-2$ km s$^{-1}$.
To clean the sample from possible contamination from binary systems, we selected only stars with a probability of being an unresolved binary smaller than 50\%, according to the criterion defined by \citet{kamann+18}.
We also subtract the average LOS velocity of the sample ($\langle v_{\rm LOS}\rangle = -18.6^{+0.2}_{-0.1}$ km s$^{-1}$) from individual velocities.
The final sample of LOS velocities used in this work thus comprises 14,601 stars.

% The catalog of LOS velocities was supplemented with public proper motion data from 
Similarly, the catalog of PM data from \citet{libralato+22} represents the most complete, homogeneous
collection of PMs of stars in the cores of stellar clusters to date.
To select stars with reliable PM estimates, we applied the quality selections described in \citet[][see their section 4]{libralato+22}, retaining 68,954 stars.
Moreover, we cleaned the sample from contaminants of the Small Magellanic Cloud taking advantage of its high velocity \citep[$\mu_{\alpha *}=-4.716\pm0.035$ mas yr$^{-1}$ and $\mu_\delta=1.325\pm0.021$ mas yr$^{-1}$,][]{anderson_king2003} relative to 47~Tucanae in proper motion space. We thus removed all the stars further than $2.6$ mas yr$^{-1}$ from the cluster bulk velocity, corresponding to more than $50$ km s$^{-1}$ \citep[i.e. larger than the central escape speed;][]{Baumgardt_Hilker_2018}.

Finally, we used the number density profile provided by \citet{deBoer_etal2019} to model the stellar density distribution of the GC. 
The authors combined Gaia DR2 data in the external regions (projected distances from the center larger than $\sim20$'), with ground-based and HST observations from \citet{Trager_etal1995} and \citet{Miocchi_etal2013}, respectively.

The dataset covers the whole cluster extent, from $\sim1$" to the cluster outskirts. In the fitting procedure, we adopted a fixed background level of $0.08$~stars~arcmin$^{-2}$ \citep[see e.g.][]{HenaultBrunet_etal2020}.

In particular, we fitted individual stellar velocities within 12" from the center. 
This distance would correspond to the radius of influence, defined by the implicit relation  $R_{\rm infl}\equiv G\,M_\bullet / \sigma^2_{\rm LOS}$ (where $\sigma_{\rm LOS} = \sigma_{\rm LOS}(R)$ is the LOS velocity dispersion, see eq.~\ref{eq:velocity_dispersion}),
of a putative IMBH with mass $M_\bullet = 10^4$ M$_\odot$, well above all previous claimed detections \citep{Kiziltan_etal2017,Mann_etal2020}, and upper limits \citep{mclaughlin_etal2006}. 
% After applying quality selections (see Section~\ref{sec:data_analysis}), 
Our final kinematic sample consists of 260 stars, with either PM or LOS velocity, and 21 stars with the full three-dimensional velocity. 
Outside 12", we used the velocity dispersion profiles computed using the same datasets.
% provided by \citet[][for the PM components]{libralato+22} and \citet[][for the LOS component]{kamann+18}.
% The kinematic sample is complemented by the stellar surface number density profile of 47 Tucanae provided by \citet{deBoer_etal2019}.

Throughout the analysis, we adopted the center reported by \citet{Goldsbury+2010}, and the kinematic distance $4.34$ kpc \citep{libralato+22}, without accounting for its $0.06$ kpc error. 
Propagation of this error on PMs would contribute at 1\% level, negligible compared to typical relative uncertainties on PM data around 16\%.

Our models are non-rotating, while there is evidence that 47~Tucanae does rotate \citep{anderson_king2003,bellini_etal2017,kamann+18}.
However, 
\citet{kamann+18} derived the dispersion profiles we used accounting for a rotationally-dependent mean velocity, and rotation is erased when deriving PMs due to local corrections. 
A residual differential rotation could be present in the LOS sample of central stars (i.e. within 12"). However, in the very central regions, the LOS rotation velocity is expected to be $\simeq 1$ km s$^{-1}$, i.e. a small fraction of the central LOS velocity dispersion \citep{kamann+18}.
We note that any residual rotation, would likely bias the model toward higher IMBH masses, as it would increase the inferred central velocity dispersion.

\section{Results}\label{sec:results}
\begin{figure*}[!th]
    \begin{minipage}{.43\textwidth}
        \centering
        \includegraphics[width=\textwidth]{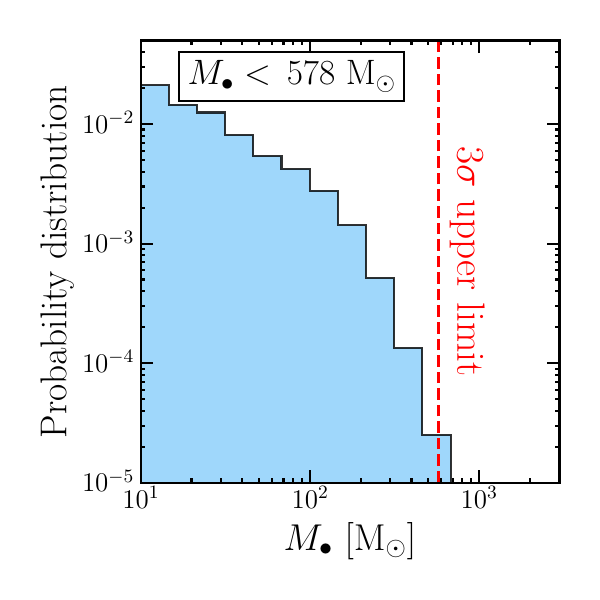}
    \end{minipage}
    \begin{minipage}{.45\textwidth}
        \centering
        \includegraphics[width=\textwidth]{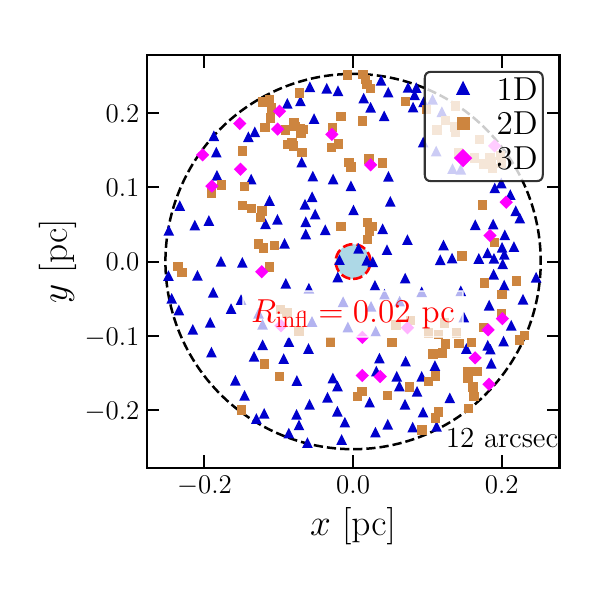}
    \end{minipage}
    \caption{
    \emph{Left panel:} posterior distribution on the BH mass (blue histogram). 
    The vertical line indicates the upper limit on the BH mass (578 M$_\odot$) containing 99.7\% ($3\sigma$) of the posterior distribution.
    \emph{Right panel:} 
    Spatial distribution of the kinematic sample of individual stars inside a circumference of radius of 12" (black curve). Each star is color-coded according to the available kinematic information: LOS velocity (i.e. 1D velocity) in dark-blue, proper motion (2D) in brown, and full kinematic information (proper motion and LOS velocity, i.e. 3D) in magenta.
    The blue shaded area indicates the region that would be influenced by a central BH with mass 578~$M_\odot$ (our $3\sigma$ upper limit), which has a radius of influence $R_{\rm infl}=0.02$~pc (red curve).
    }
    \label{fig:posterior_blackhole}
\end{figure*}
\begin{figure*}[!th]
    \centering
    \includegraphics[width=\textwidth]{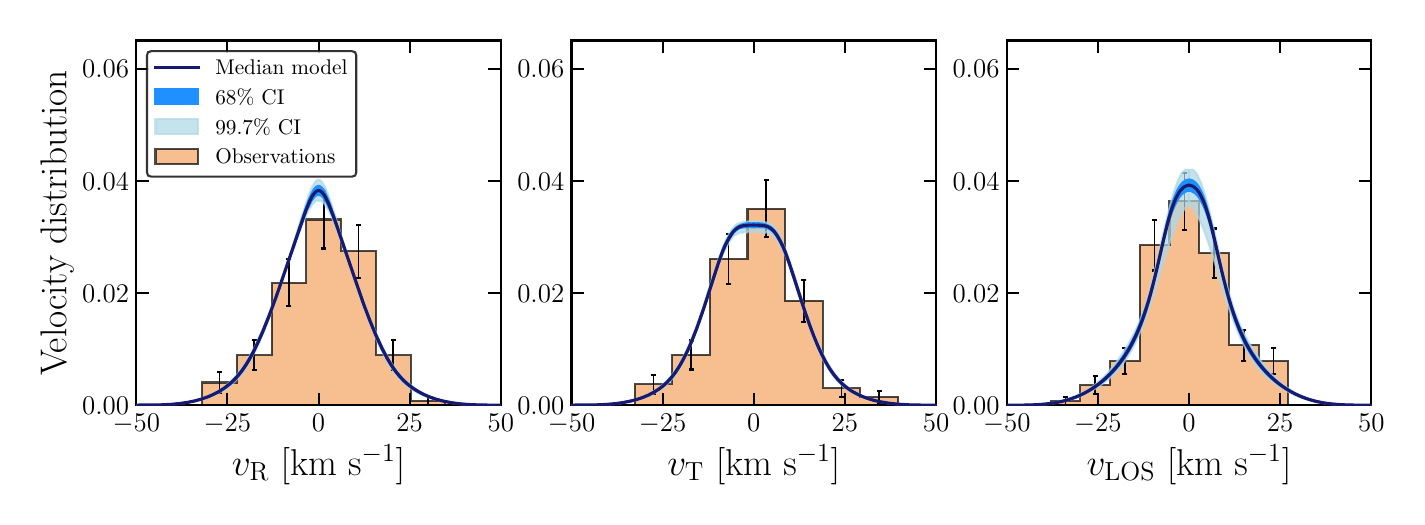}
    \caption{
    % \cn{RIGHTMOST PANEL: perche' la distribuzione e' piu' peaked rispetto agli altri due pannelli (almeno nel modello)? e' dovuto al fatto che gli errori LOS sono piu' piccoli? ad altro? se lo capiamo... sarebbe una cosa da dire nel testo.} 
    Observed velocity distribution within 12" from the cluster center, along the radial (leftmost panel), tangential (central panel), and LOS (rightmost panel) directions. 
    The bars indicate uncertainties estimated as Poissonian errors on the bin counts.
    The median models (solid line), 68\% ($1\sigma$), and 99.7\% ($3\sigma$) CIs (shaded areas) are shown in blue (see text for details). 
    % The velocity distributions computed from the model were convolved with a Gaussian, with a standard deviation equal to the median error on observed velocities, and were integrated over the radial extension covered by the data, to allow for a quantitative comparison.
    % We further emphasize though that histograms and related errors are shown for visualization purposes only, as the model does not fit the binned distributions, but is compared with individual star data.
    }
    \label{fig:velocitydistribution_12arcsec}
\end{figure*}
\begin{figure*}[!th]
    \centering
    \includegraphics[width=\textwidth]{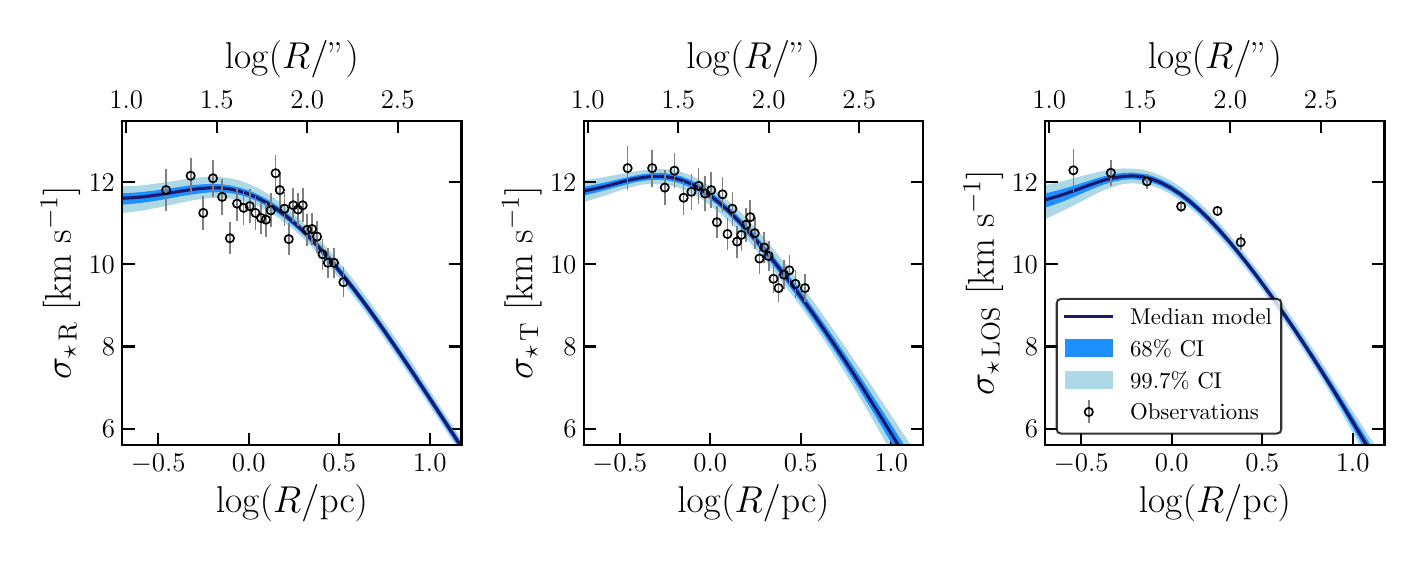}
    \caption{Projected velocity dispersion profiles along the radial (left panel), tangential (central panel), and LOS (right panel) directions. Observations are shown as black points along with $1\sigma$ error bars. The blue line is the median model, while the shaded areas represent the 68\% and the 99.7\% CIs.}
    \label{fig:velocity_dispersion}
\end{figure*}
\begin{figure}[!th]
    \centering
    \includegraphics[width=.45\textwidth]{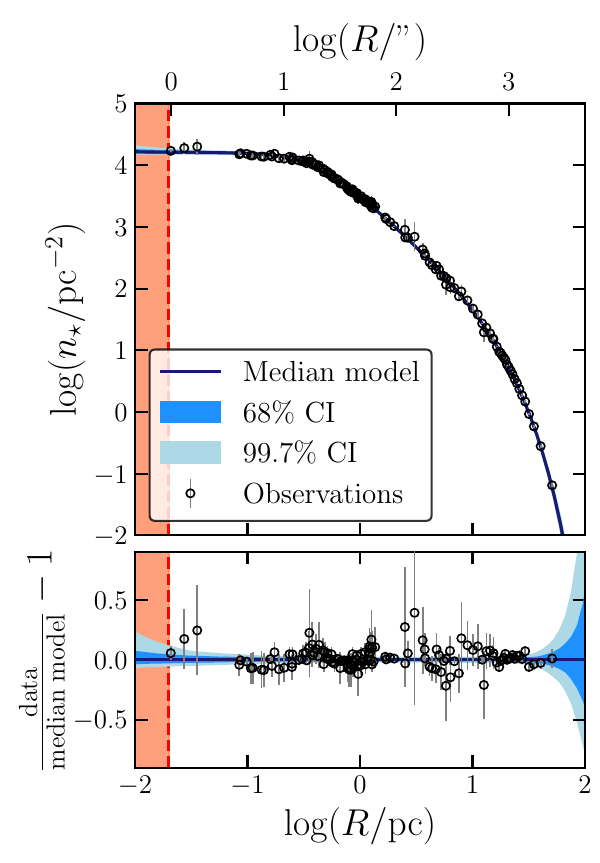}
    \caption{Surface number density profile as a function of the projected distance from the cluster center. Observations are shown as black points (with $1\sigma$ error bars) whereas the model is shown in blue. The 68\% and 99.7\% CIs are also shown as shaded areas.
    The vertical dashed line represents the $3\sigma$ upper limit (0.02 pc) on the BH $R_{\rm infl}$.
    % \cn{Notazione: nell'appendice $\Sigma_\star$ e' definita come mass density, non number density. In questa figura potremmo usare $n_\star$ per la number density. Da qualche parte (ad esempio in appendice A) dovremmo scrivere esplicitamente come, per i modelli, otteniamo $n_\star$ da $\Sigma_\star$. mi sembra di capire che $n_\star=m \Sigma_\star$, dove $m$ e' uno dei parametri liberi (by the way, se e' cosi', $m$ in table C1 non può essere adimensionale. }
    }
    \label{fig:density_profile}
\end{figure}
\begin{figure}[!th]
    \centering
    \includegraphics[width=.5\textwidth]{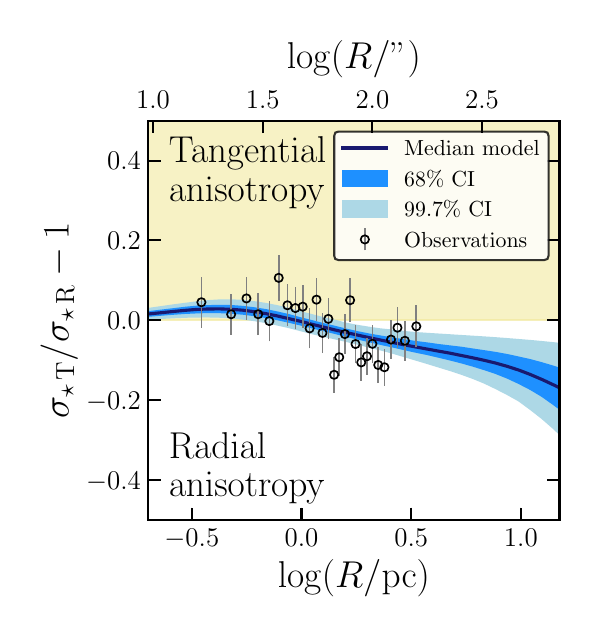}
    \caption{Projected velocity anisotropy profiles. 
    Positive values of the anisotropy parameters correspond to tangential anisotropy, whereas negative ones to radial anisotropy.
    The median model is shown in blue and the corresponding CIs are shown as shaded areas.
    The black points are observational data from \citet{libralato+22} with $1\sigma$ error bars.
    }
    \label{fig:anisotropy}
\end{figure}

The left panel of Fig.~\ref{fig:posterior_blackhole} shows the posterior distribution of the IMBH mass.
According to our analysis, we find no evidence of an IMBH in 47~Tucanae. We rather put an upper limit of $M_\bullet<578$ M$_\odot$ at the $3\sigma$ level. 
This is the most stringent upper limit on the mass of a putative central dark component in 47~Tucanae ever achieved by any dynamical study.
The right panel of Fig.~\ref{fig:posterior_blackhole} shows the $3\sigma$ upper limit on the IMBH $R_{\rm infl}$, overplotted to the on-sky distribution of stars closer than 12" to the center.
It is clear that the kinematics of these stars put a very tight constraint on $R_{\rm infl}$, 
whose upper limit is comparable to the distance from the center of the innermost stars.
We further verified this point, by performing additional fits where first the individual stars inside 12", and then also the velocity dispersion profiles were removed. We found that the upper limit on the IMBH mass increases to a few thousand and to several hundred thousand solar masses respectively. 

Fig.~\ref{fig:velocitydistribution_12arcsec} shows the PM and LOS velocity distributions for stars within 12" from the center. Overplotted to the observations, we show the median model, and the 68\% and 99.7\% credible intervals (CIs) for the corresponding velocity distributions. 
Each model was 
convolved with a Gaussian distribution with a standard deviation equal to the observational median error in each component, and was integrated over the radial extension covered by the datasets.
The model reproduces the observed velocity distributions up to the tails (Fig.~\ref{fig:velocitydistribution_12arcsec}). We emphasize though that we did not fit the binned histograms, whereas we used an individual-star approach fully exploiting the datasets (see eq.~\ref{eq:likelihood_singlestars}). 

The very good agreement between the model and the data can be 
further observed in
Fig.s~\ref{fig:velocity_dispersion}, and \ref{fig:density_profile}, which show the projected velocity dispersion profiles and the stellar density profile, compared with the median and CIs of the corresponding theoretical profiles. 
We also compared (Fig.~\ref{fig:anisotropy}) our model with measurements of the projected velocity anisotropy \citep[data from][]{libralato+22}, defined as $\sigma_{\star\,{\rm T}}/\sigma_{\star\,{\rm R}} - 1$, with $\sigma_{\star\,{\rm R}}$ and $\sigma_{\star\,{\rm T}}$ the radial and tangential velocity dispersion components, respectively (see eq.~\ref{eq:velocity_dispersion}).
Figure~\ref{fig:anisotropy} shows that our model can reproduce the system velocity anisotropy remarkably well also compared to previous studies \citep[see e.g.][]{Dickson_etal2023}.
% Consistent with observations, the system is isotropic in the center and mildly radially anisotropic in the external regions.

\section{Comparison with previous works} \label{sec:comparison_previous_works}
\begin{figure}
    \centering
    \includegraphics[width=.5\textwidth]{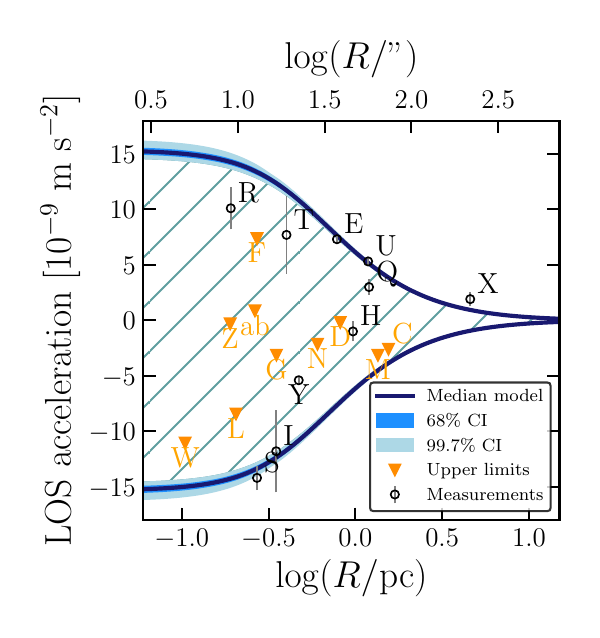}
    \caption{LOS acceleration as a function of the projected distance from the center of 47~Tucanae. 
    Black points are measurements, while orange ones are upper limits. Both were obtained from pulsars \citep[data from][]{ridolfi_etal2016,Freire_etal2017}. The label of each pulsar is shown either in orange or black.
    The blue line shows the maximum (if positive) and minimum (if negative) LOS acceleration allowed by our dynamical model at a given distance from the center. 
    Similarly, the 68\% and 99.7\% CIs on the maximum and minimum LOS acceleration are shown as shaded areas.
    The hatched area is the allowed LOS acceleration space.}
    \label{fig:psr_acceleration}
\end{figure}
Our result is in tension with some previous studies claiming the presence of a massive IMBH in 47~Tucanae \citep[see e.g.][]{Kiziltan_etal2017,Mann_etal2020}. In this section, we delve into the possible reasons for discrepancies.

\citet{Kiziltan_etal2017} used spin-down measurements for nineteen MSPs identified in 47~Tucanae.
Comparing acceleration data with $N$-body simulations, they found evidence for a massive central BH with mass $M_\bullet=2300^{+1500}_{-850}$ M$_\odot$.
While a direct comparison with the study of \citet{Kiziltan_etal2017} is not straightforward, as different dynamical models and data were used, we note that it is in general hard to perform a large exploration of possible initial conditions using $N$-body simulations.
Dynamical models of equilibrium, on the other hand, allow us to perform a systematic exploration of the parameter space. 
In addition, in our model-data comparison, we fit simultaneously the spatial distribution and the full velocity distribution using individual stars,
while \citet{Kiziltan_etal2017} analyzed only those
$N$-body simulations that better reproduced
the density profile and the LOS velocity dispersion, which may not be representative of the full cluster kinematics.

However, as a further check, we verified whether our best-fit model is able to reproduce the measurements of MSPs.
Fig.~\ref{fig:psr_acceleration} shows the cluster LOS acceleration data from the MSP sample \citep[data from][and \citealt{Freire_etal2017}]{ridolfi_etal2016} used by \citet{Kiziltan_etal2017}, as well as the maximum and minimum LOS acceleration allowed by our dynamical model.
This quantity was computed as the maximum projection along the LOS of the radial acceleration, $a(r)$, at any given projected distance from the center, $R$, i.e. ${\rm max}\Big( a(r)\,\sqrt{1-(R/r)^2} \Big)\,\forall r\geq R$. 
Interestingly, our model is compatible with all the upper limits and also with the central pulsars showing the highest accelerations (such as 47~Tuc-E, 47~Tuc-U, 47~Tuc-I, and 47~Tuc-S) without the need for an IMBH more massive than $578$~M$_\odot$ (at the $3\sigma$ level).
The only outlier might be 47~Tuc-X, still compatible within $2\sigma$ with the median model.

Using dynamical models based on the Jeans equations coupled with PM data of the cluster center from HST, \citet{Mann_etal2020} found that a massive IMBH with mass $808-4610$ M$_\odot$ is required to explain the central kinematics.
While employing the same dataset would be the best approach 
to understand the possible reasons that lead to a discrepancy, we note that \citet{Mann_etal2020}
only attempted to reproduce the system velocity dispersion. 
Our DF-based models and individual-star approach (see Appendix~\ref{sec:posterior_definition}), on the contrary, make the most of the kinematic sample since it does not condensate the kinematic information in few radial bins, but rather models, in a continuous way, the full shape of the cluster's velocity distribution in the center.
This approach provides more stringent constraints on the presence of a putative massive dark component in the center \citep[see][]{Pascale_etal2019}.
Furthermore, we also used LOS data to probe the 3D kinematics.

Finally, we note that our result is consistent with findings by 
\citet{mclaughlin_etal2006} and \citet{HenaultBrunet_etal2020}. 
Using HST data and Jeans modeling, \citet{mclaughlin_etal2006} put an upper limit of about $1578$ M$_\odot$ at the $1\sigma$ level, compatible with the much more stringent upper limit set by this study (578 M$_\odot$ at $3\sigma$).
Moreover, \citet{HenaultBrunet_etal2020} constrained the overall mass budget in dark remnants (stellar-mass BH, neutron stars, and white dwarfs) possibly harbored at the center of 47~Tucanae. The upper limit set by the current work is consistent with the mass budget of $M_{\rm remnant} = 430^{+386}_{-301}$ M$_\odot$ found by \citet{HenaultBrunet_etal2020}. 
Although in our analysis we considered only a point-like central mass (like an IMBH), the upper limit on the central mass we found should apply also to an extended spherical central object.

\section{Conclusions}
\label{sec:conclusions}
In this work we addressed the problem of the presence of a putative IMBH at the center of the GC  47~Tucanae, using dynamical models based on DFs depending on the action integrals.
We modeled state-of-the-art data providing information on the spatial distribution, and both LOS and on-sky kinematics up to the very central regions of the cluster. 
Also, we employed a star-by-star approach in the central region to fully exploit the data and model the full shape of the velocity distribution.
According to our analysis, we rule out (at the $3\sigma$ level) the presence of a dark central component more massive than $578$ M$_\odot$. To date, this is the most stringent upper limit that has been set in 47~Tucanae by any dynamical study.
While consistent with other studies \citep[e.g.][]{mclaughlin_etal2006,HenaultBrunet_etal2020},
our result is in tension with those studies claiming the detection of an IMBH in 47~Tucanae \citep[see e.g.][]{Mann_etal2020,Kiziltan_etal2017}.
% In Appendix~\ref{sec:comparison_previous_works} we delve into the possible reasons for the discrepancy.

Despite the very stringent upper limit we put in this study, more sophisticated dynamical models and novel, precise data would shed further light on the nature of a putative central dark component in 47~Tucanae as well as in other GCs.
For instance, multi-mass modeling would allow us to account for stellar evolution and mass segregation \citep{GielesZocchi_2015}. 

From the data point of view, larger coverage of the central regions and progressively better-defined cluster centers and distances would certainly provide more robust results. 
Future facilities, such as the Extremely Large Telescope, will measure the stellar kinematics of GC centers with unprecedented accuracy, likely providing new exciting data on the topic.

Finally, we note that the methodology presented in this work could be applied to any GC in our Galaxy, regardless of the particular data available either PMs or LOS velocities only, or full 3D kinematic information as done in the present study.
In addition, it is not limited to GCs but it could be also used to explore the presence of central BHs in external galaxies (Pascale et al.\ in preparation).

\begin{acknowledgements}
We thank the anonymous referee for their comments.
The authors are grateful to Sebastian Kamann for providing us with the MUSE data and to Andrea Bellini for fruitful discussions.
\end{acknowledgements}

\bibliographystyle{aa}
\bibliography{references}

\begin{thebibliography}{45}
\expandafter\ifx\csname natexlab\endcsname\relax\def\natexlab#1{#1}\fi

\bibitem[{{Abbate} {et~al.}(2018){Abbate}, {Possenti}, {Ridolfi}, {Freire},
  {Camilo}, {Manchester}, \& {D'Amico}}]{Abbate_etal2018}
{Abbate}, F., {Possenti}, A., {Ridolfi}, A., {et~al.} 2018, \mnras, 481, 627

\bibitem[{{Abbott} {et~al.}(2020){Abbott}, {Abbott}, {Abraham}, {Acernese},
  {Ackley}, {Adams}, {Adhikari}, {Adya}, {Affeldt}, {Agathos}, {Agatsuma},
  {Aggarwal}, {Aguiar}, {Aich}, {Aiello}, {Ain}, {Ajith}, {Akcay}, {Allen},
  {Allocca}, {Altin}, {Amato}, {Anand}, {Ananyeva}, {Anderson}, {Anderson},
  {Angelova}, {Ansoldi}, {Antier}, {Appert}, {Arai}, {Araya}, {Areeda},
  {Ar{\`e}ne}, {Arnaud}, {Aronson}, {Arun}, {Asali}, {Ascenzi}, {Ashton},
  {Aston}, {Astone}, {Aubin}, {Aufmuth}, {AultONeal}, {Austin}, {Avendano},
  {Babak}, {Bacon}, {Badaracco}, {Bader}, {Bae}, {Baer}, {Baird}, {Baldaccini},
  {Ballardin}, {Ballmer}, {Bals}, {Balsamo}, {Baltus}, {Banagiri}, {Bankar},
  {Bankar}, {Barayoga}, {Barbieri}, {Barish}, {Barker}, {Barkett}, {Barneo},
  {Barone}, {Barr}, {Barsotti}, {Barsuglia}, {Barta}, {Bartlett}, {Bartos},
  {Bassiri}, {Basti}, {Bawaj}, {Bayley}, {Bazzan}, {B{\'e}csy}, {Bejger},
  {Belahcene}, {Bell}, {Beniwal}, {Benjamin}, {Bentley}, {Bergamin}, {Berger},
  {Bergmann}, {Bernuzzi}, {Berry}, {Bersanetti}, {Bertolini}, {Betzwieser},
  {Bhandare}, {Bhandari}, {Bidler}, {Biggs}, {Bilenko}, {Billingsley},
  {Birney}, {Birnholtz}, {Biscans}, {Bischi}, {Biscoveanu}, {Bisht},
  {Bissenbayeva}, {Bitossi}, {Bizouard}, {Blackburn}, {Blackman}, {Blair},
  {Blair}, {Blair}, {Bobba}, {Bode}, {Boer}, {Boetzel}, {Bogaert}, {Bondu},
  {Bonilla}, {Bonnand}, {Booker}, {Boom}, {Bork}, {Boschi}, {Bose},
  {Bossilkov}, {Bosveld}, {Bouffanais}, {Bozzi}, {Bradaschia}, {Brady},
  {Bramley}, {Branchesi}, {Brau}, {Breschi}, {Briant}, {Briggs}, {Brighenti},
  {Brillet}, {Brinkmann}, {Brockill}, {Brooks}, {Brooks}, {Brown}, {Brunett},
  {Bruno}, {Bruntz}, {Buikema}, {Bulik}, {Bulten}, {Buonanno}, {Buscicchio},
  {Buskulic}, {Byer}, {Cabero}, {Cadonati}, {Cagnoli}, {Cahillane},
  {Calder{\'o}n Bustillo}, {Callaghan}, {Callister}, {Calloni}, {Camp},
  {Canepa}, {Cannon}, {Cao}, {Cao}, {Carapella}, {Carbognani}, {Caride},
  {Carney}, {Carullo}, {Casanueva Diaz}, {Casentini}, {Casta{\~n}eda},
  {Caudill}, {Cavagli{\`a}}, {Cavalier}, {Cavalieri}, {Cella},
  {Cerd{\'a}-Dur{\'a}n}, {Cesarini}, {Chaibi}, {Chakravarti}, {Chan}, {Chan},
  {Chandra}, {Chao}, {Charlton}, {Chase}, {Chassande-Mottin}, {Chatterjee},
  {Chaturvedi}, {Chatziioannou}, {Chen}, {Chen}, {Chen}, {Cheng}, {Cheong},
  {Chia}, {Chiadini}, {Chierici}, {Chincarini}, {Chiummo}, {Cho}, {Cho}, {Cho},
  {Christensen}, {Chu}, {Chua}, {Chung}, {Chung}, {Ciani}, {Ciecielag},
  {Cie{\'s}lar}, {Ciobanu}, {Ciolfi}, {Cipriano}, {Cirone}, {Clara}, {Clark},
  {Clearwater}, {Clesse}, {Cleva}, {Coccia}, {Cohadon}, {Cohen}, {Colleoni},
  {Collette}, {Collins}, {Colpi}, {Constancio}, {Conti}, {Cooper}, {Corban},
  {Corbitt}, {Cordero-Carri{\'o}n}, {Corezzi}, {Corley}, {Cornish}, {Corre},
  {Corsi}, {Cortese}, {Costa}, {Cotesta}, {Coughlin}, {Coughlin}, {Coulon},
  {Countryman}, {Couvares}, {Covas}, {Coward}, {Cowart}, {Coyne}, {Coyne},
  {Creighton}, {Creighton}, {Cripe}, {Croquette}, {Crowder}, {Cudell},
  {Cullen}, {Cumming}, {Cummings}, {Cunningham}, {Cuoco}, {Curylo}, {Canton},
  {D{\'a}lya}, {Dana}, {Daneshgaran-Bajastani}, {D'Angelo}, {Danilishin},
  {D'Antonio}, {Danzmann}, {Darsow-Fromm}, {Dasgupta}, {Datrier}, {Dattilo},
  {Dave}, {Davier}, {Davies}, {Davis}, {Daw}, {DeBra}, {Deenadayalan},
  {Degallaix}, {De Laurentis}, {Del{\'e}glise}, {Delfavero}, {De Lillo}, {Del
  Pozzo}, {DeMarchi}, {D'Emilio}, {Demos}, {Dent}, {De Pietri}, {De Rosa}, {De
  Rossi}, {DeSalvo}, {de Varona}, {Dhurandhar}, {D{\'\i}az}, {Diaz-Ortiz},
  {Dietrich}, {Di Fiore}, {Di Fronzo}, {Di Giorgio}, {Di Giovanni}, {Di
  Giovanni}, {Di Girolamo}, {Di Lieto}, {Ding}, {Di Pace}, {Di Palma}, {Di
  Renzo}, {Divakarla}, {Dmitriev}, {Doctor}, {Donovan}, {Dooley}, {Doravari},
  {Dorrington}, {Downes}, {Drago}, {Driggers}, {Du}, {Ducoin}, {Dupej},
  {Durante}, {D'Urso}, {Dwyer}, {Easter}, {Eddolls}, {Edelman}, {Edo}, {Edy},
  {Effler}, {Ehrens}, {Eichholz}, {Eikenberry}, {Eisenmann}, {Eisenstein},
  {Ejlli}, {Errico}, {Essick}, {Estelles}, {Estevez}, {Etienne}, {Etzel},
  {Evans}, {Evans}, {Ewing}, {Fafone}, {Fairhurst}, {Fan}, {Farinon}, {Farr},
  {Farr}, {Fauchon-Jones}, {Favata}, {Fays}, {Fazio}, {Feicht}, {Fejer},
  {Feng}, {Fenyvesi}, {Ferguson}, {Fernandez-Galiana}, {Ferrante}, {Ferreira},
  {Ferreira}, {Fidecaro}, {Fiori}, {Fiorucci}, {Fishbach}, {Fisher},
  {Fittipaldi}, {Fitz-Axen}, {Fiumara}, {Flaminio}, {Floden}, {Flynn}, {Fong},
  {Font}, {Forsyth}, {Fournier}, {Frasca}, {Frasconi}, {Frei}, {Freise},
  {Frey}, {Frey}, {Fritschel}, {Frolov}, {Fronz{\`e}}, {Fulda}, {Fyffe},
  {Gabbard}, {Gadre}, {Gaebel}, {Gair}, {Galaudage}, {Ganapathy}, {Ganguly},
  {Gaonkar}, {Garc{\'\i}a-Quir{\'o}s}, {Garufi}, {Gateley}, {Gaudio},
  {Gayathri}, {Gemme}, {Genin}, {Gennai}, {George}, {George}, {Gergely},
  {Ghonge}, {Ghosh}, {Ghosh}, {Ghosh}, {Giacomazzo}, {Giaime}, {Giardina},
  {Gibson}, {Gier}, {Gill}, {Glanzer}, {Gniesmer}, {Godwin}, {Goetz}, {Goetz},
  {Gohlke}, {Goncharov}, {Gonz{\'a}lez}, {Gopakumar}, {Gossan}, {Gosselin},
  {Gouaty}, {Grace}, {Grado}, {Granata}, {Grant}, {Gras}, {Grassia}, {Gray},
  {Gray}, {Greco}, {Green}, {Green}, {Gretarsson}, {Griggs}, {Grignani},
  {Grimaldi}, {Grimm}, {Grote}, {Grunewald}, {Gruning}, {Guidi}, {Guimaraes},
  {Guix{\'e}}, {Gulati}, {Guo}, {Gupta}, {Gupta}, {Gupta}, {Gustafson},
  {Gustafson}, {Haegel}, {Halim}, {Hall}, {Hamilton}, {Hammond}, {Haney},
  {Hanke}, {Hanks}, {Hanna}, {Hannam}, {Hannuksela}, {Hansen}, {Hanson},
  {Harder}, {Hardwick}, {Haris}, {Harms}, {Harry}, {Harry}, {Hasskew},
  {Haster}, {Haughian}, {Hayes}, {Healy}, {Heidmann}, {Heintze}, {Heinze},
  {Heitmann}, {Hellman}, {Hello}, {Hemming}, {Hendry}, {Heng}, {Hennes},
  {Hennig}, {Heurs}, {Hild}, {Hinderer}, {Hoback}, {Hochheim}, {Hofgard},
  {Hofman}, {Holgado}, {Holland}, {Holt}, {Holz}, {Hopkins}, {Horst}, {Hough},
  {Howell}, {Hoy}, {Huang}, {H{\"u}bner}, {Huerta}, {Huet}, {Hughey}, {Hui},
  {Husa}, {Huttner}, {Huxford}, {Huynh-Dinh}, {Idzkowski}, {Iess}, {Inchauspe},
  {Ingram}, {Intini}, {Isac}, {Isi}, {Iyer}, {Jacqmin}, {Jadhav}, {Jadhav},
  {James}, {Jani}, {Janthalur}, {Jaranowski}, {Jariwala}, {Jaume}, {Jenkins},
  {Jiang}, {Johns}, {Johnson-McDaniel}, {Jones}, {Jones}, {Jones}, {Jones},
  {Jones}, {Jonker}, {Ju}, {Junker}, {Kalaghatgi}, {Kalogera}, {Kamai},
  {Kandhasamy}, {Kang}, {Kanner}, {Kapadia}, {Karki}, {Kashyap}, {Kasprzack},
  {Kastaun}, {Katsanevas}, {Katsavounidis}, {Katzman}, {Kaufer}, {Kawabe},
  {K{\'e}f{\'e}lian}, {Keitel}, {Keivani}, {Kennedy}, {Key}, {Khadka},
  {Khalili}, {Khan}, {Khan}, {Khan}, {Khazanov}, {Khetan}, {Khursheed},
  {Kijbunchoo}, {Kim}, {Kim}, {Kim}, {Kim}, {Kim}, {Kim}, {Kim}, {Kimball},
  {King}, {Kinley-Hanlon}, {Kirchhoff}, {Kissel}, {Kleybolte}, {Klimenko},
  {Knowles}, {Knyazev}, {Koch}, {Koehlenbeck}, {Koekoek}, {Koley},
  {Kondrashov}, {Kontos}, {Koper}, {Korobko}, {Korth}, {Kovalam}, {Kozak},
  {Kringel}, {Krishnendu}, {Kr{\'o}lak}, {Krupinski}, {Kuehn}, {Kumar},
  {Kumar}, {Kumar}, {Kumar}, {Kumar}, {Kuo}, {Kutynia}, {Lackey}, {Laghi},
  {Lalande}, {Lam}, {Lamberts}, {Landry}, {Lane}, {Lang}, {Lange}, {Lantz},
  {Lanza}, {La Rosa}, {Lartaux-Vollard}, {Lasky}, {Laxen}, {Lazzarini},
  {Lazzaro}, {Leaci}, {Leavey}, {Lecoeuche}, {Lee}, {Lee}, {Lee}, {Lee}, {Lee},
  {Lehmann}, {Leroy}, {Letendre}, {Levin}, {Li}, {Li}, {li}, {Li}, {Li},
  {Linde}, {Linker}, {Linley}, {Littenberg}, {Liu}, {Liu},
  {Llorens-Monteagudo}, {Lo}, {Lockwood}, {London}, {Longo}, {Lorenzini},
  {Loriette}, {Lormand}, {Losurdo}, {Lough}, {Lousto}, {Lovelace}, {L{\"u}ck},
  {Lumaca}, {Lundgren}, {Ma}, {Macas}, {Macfoy}, {MacInnis}, {Macleod},
  {MacMillan}, {Macquet}, {Maga{\~n}a Hernandez}, {Maga{\~n}a-Sandoval},
  {Magee}, {Majorana}, {Maksimovic}, {Malik}, {Man}, {Mandic}, {Mangano},
  {Mansell}, {Manske}, {Mantovani}, {Mapelli}, {Marchesoni}, {Marion},
  {M{\'a}rka}, {M{\'a}rka}, {Markakis}, {Markosyan}, {Markowitz}, {Maros},
  {Marquina}, {Marsat}, {Martelli}, {Martin}, {Martin}, {Martinez}, {Martynov},
  {Masalehdan}, {Mason}, {Massera}, {Masserot}, {Massinger}, {Masso-Reid},
  {Mastrogiovanni}, {Matas}, {Matichard}, {Mavalvala}, {Maynard}, {McCann},
  {McCarthy}, {McClelland}, {McCormick}, {McCuller}, {McGuire}, {McIsaac},
  {McIver}, {McManus}, {McRae}, {McWilliams}, {Meacher}, {Meadors}, {Mehmet},
  {Mehta}, {Mejuto Villa}, {Melatos}, {Mendell}, {Mercer}, {Mereni}, {Merfeld},
  {Merilh}, {Merritt}, {Merzougui}, {Meshkov}, {Messenger}, {Messick},
  {Metzdorff}, {Meyers}, {Meylahn}, {Mhaske}, {Miani}, {Miao}, {Michaloliakos},
  {Michel}, {Middleton}, {Milano}, {Miller}, {Millhouse}, {Mills}, {Milotti},
  {Milovich-Goff}, {Minazzoli}, {Minenkov}, {Mishkin}, {Mishra}, {Mistry},
  {Mitra}, {Mitrofanov}, {Mitselmakher}, {Mittleman}, {Mo}, {Mogushi},
  {Mohapatra}, {Mohite}, {Molina-Ruiz}, {Mondin}, {Montani}, {Moore}, {Moraru},
  {Morawski}, {Moreno}, {Morisaki}, {Mours}, {Mow-Lowry}, {Mozzon},
  {Muciaccia}, {Mukherjee}, {Mukherjee}, {Mukherjee}, {Mukherjee}, {Mukund},
  {Mullavey}, {Munch}, {Mu{\~n}iz}, {Murray}, {Nagar}, {Nardecchia},
  {Naticchioni}, {Nayak}, {Neil}, {Neilson}, {Nelemans}, {Nelson}, {Nery},
  {Neunzert}, {Ng}, {Ng}, {Nguyen}, {Nguyen}, {Nichols}, {Nichols}, {Nissanke},
  {Nitz}, {Nocera}, {Noh}, {North}, {Nothard}, {Nuttall}, {Oberling},
  {O'Brien}, {Oganesyan}, {Ogin}, {Oh}, {Oh}, {Ohme}, {Ohta}, {Okada},
  {Oliver}, {Olivetto}, {Oppermann}, {Oram}, {O'Reilly}, {Ormiston}, {Ortega},
  {O'Shaughnessy}, {Ossokine}, {Osthelder}, {Ottaway}, {Overmier}, {Owen},
  {Pace}, {Pagano}, {Page}, {Pagliaroli}, {Pai}, {Pai}, {Palamos}, {Palashov},
  {Palomba}, {Pan}, {Panda}, {Pang}, {Pankow}, {Pannarale}, {Pant}, {Paoletti},
  {Paoli}, {Parida}, {Parker}, {Pascucci}, {Pasqualetti}, {Passaquieti},
  {Passuello}, {Patricelli}, {Payne}, {Pearlstone}, {Pechsiri}, {Pedersen},
  {Pedraza}, {Pele}, {Penn}, {Perego}, {Perez}, {P{\'e}rigois}, {Perreca},
  {Perri{\`e}s}, {Petermann}, {Pfeiffer}, {Phelps}, {Phukon}, {Piccinni},
  {Pichot}, {Piendibene}, {Piergiovanni}, {Pierro}, {Pillant}, {Pinard},
  {Pinto}, {Piotrzkowski}, {Pirello}, {Pitkin}, {Plastino}, {Poggiani}, {Pong},
  {Ponrathnam}, {Popolizio}, {Porter}, {Powell}, {Prajapati}, {Prasai},
  {Prasanna}, {Pratten}, {Prestegard}, {Principe}, {Prodi}, {Prokhorov},
  {Punturo}, {Puppo}, {P{\"u}rrer}, {Qi}, {Quetschke}, {Quinonez}, {Raab},
  {Raaijmakers}, {Radkins}, {Radulesco}, {Raffai}, {Rafferty}, {Raja}, {Rajan},
  {Rajbhandari}, {Rakhmanov}, {Ramirez}, {Ramos-Buades}, {Rana}, {Rao},
  {Rapagnani}, {Raymond}, {Razzano}, {Read}, {Regimbau}, {Rei}, {Reid},
  {Reitze}, {Rettegno}, {Ricci}, {Richardson}, {Richardson}, {Ricker},
  {Riemenschneider}, {Riles}, {Rizzo}, {Robertson}, {Robinet}, {Rocchi},
  {Rodriguez-Soto}, {Rolland}, {Rollins}, {Roma}, {Romanelli}, {Romano},
  {Romel}, {Romero-Shaw}, {Romie}, {Rose}, {Rose}, {Rose}, {Rosi{\'n}ska},
  {Rosofsky}, {Ross}, {Rowan}, {Rowlinson}, {Roy}, {Roy}, {Roy}, {Ruggi},
  {Rutins}, {Ryan}, {Sachdev}, {Sadecki}, {Sakellariadou}, {Salafia},
  {Salconi}, {Saleem}, {Salemi}, {Samajdar}, {Sanchez}, {Sanchez},
  {Sanchis-Gual}, {Sanders}, {Santiago}, {Santos}, {Sarin}, {Sassolas},
  {Sathyaprakash}, {Sauter}, {Savage}, {Savant}, {Sawant}, {Sayah}, {Schaetzl},
  {Schale}, {Scheel}, {Scheuer}, {Schmidt}, {Schnabel}, {Schofield},
  {Sch{\"o}nbeck}, {Schreiber}, {Schulte}, {Schutz}, {Schwarm}, {Schwartz},
  {Scott}, {Scott}, {Seidel}, {Sellers}, {Sengupta}, {Sennett}, {Sentenac},
  {Sequino}, {Sergeev}, {Setyawati}, {Shaddock}, {Shaffer}, {Sharifi},
  {Shahriar}, {Sharma}, {Sharma}, {Shawhan}, {Shen}, {Shikauchi}, {Shink},
  {Shoemaker}, {Shoemaker}, {Shukla}, {ShyamSundar}, {Siellez}, {Sieniawska},
  {Sigg}, {Singer}, {Singh}, {Singh}, {Singha}, {Singhal}, {Sintes}, {Sipala},
  {Skliris}, {Slagmolen}, {Slaven-Blair}, {Smetana}, {Smith}, {Smith},
  {Somala}, {Son}, {Soni}, {Sorazu}, {Sordini}, {Sorrentino}, {Souradeep},
  {Sowell}, {Spencer}, {Spera}, {Srivastava}, {Srivastava}, {Staats},
  {Stachie}, {Standke}, {Steer}, {Steinke}, {Steinlechner}, {Steinlechner},
  {Steinmeyer}, {Stevenson}, {Stocks}, {Stops}, {Stover}, {Strain}, {Stratta},
  {Strunk}, {Sturani}, {Stuver}, {Sudhagar}, {Sudhir}, {Summerscales}, {Sun},
  {Sunil}, {Sur}, {Suresh}, {Sutton}, {Swinkels}, {Szczepa{\'n}czyk}, {Tacca},
  {Tait}, {Talbot}, {Tanasijczuk}, {Tanner}, {Tao}, {T{\'a}pai}, {Tapia},
  {Tapia San Martin}, {Tasson}, {Taylor}, {Tenorio}, {Terkowski},
  {Thirugnanasambandam}, {Thomas}, {Thomas}, {Thompson}, {Thondapu}, {Thorne},
  {Thrane}, {Tinsman}, {Saravanan}, {Tiwari}, {Tiwari}, {Tiwari}, {Toland},
  {Tonelli}, {Tornasi}, {Torres-Forn{\'e}}, {Torrie}, {Tosta e Melo},
  {T{\"o}yr{\"a}}, {Travasso}, {Traylor}, {Tringali}, {Tripathee}, {Trovato},
  {Trudeau}, {Tsang}, {Tse}, {Tso}, {Tsukada}, {Tsuna}, {Tsutsui}, {Turconi},
  {Ubhi}, {Udall}, {Ueno}, {Ugolini}, {Unnikrishnan}, {Urban}, {Usman},
  {Utina}, {Vahlbruch}, {Vajente}, {Valdes}, {Valentini}, {van Bakel}, {van
  Beuzekom}, {van den Brand}, {Van Den Broeck}, {Vander-Hyde}, {van der
  Schaaf}, {Van Heijningen}, {van Veggel}, {Vardaro}, {Varma}, {Vass},
  {Vas{\'u}th}, {Vecchio}, {Vedovato}, {Veitch}, {Veitch}, {Venkateswara},
  {Venugopalan}, {Verkindt}, {Veske}, {Vetrano}, {Vicer{\'e}}, {Viets},
  {Vinciguerra}, {Vine}, {Vinet}, {Vitale}, {Vivanco}, {Vo}, {Vocca},
  {Vorvick}, {Vyatchanin}, {Wade}, {Wade}, {Wade}, {Walet}, {Walker},
  {Wallace}, {Wallace}, {Walsh}, {Wang}, {Wang}, {Wang}, {Ward}, {Warden},
  {Warner}, {Was}, {Watchi}, {Weaver}, {Wei}, {Weinert}, {Weinstein}, {Weiss},
  {Wellmann}, {Wen}, {We{\ss}els}, {Westhouse}, {Wette}, {Whelan}, {Whiting},
  {Whittle}, {Wilken}, {Williams}, {Willis}, {Willke}, {Winkler}, {Wipf},
  {Wittel}, {Woan}, {Woehler}, {Wofford}, {Wong}, {Wright}, {Wu}, {Wysocki},
  {Xiao}, {Yamamoto}, {Yang}, {Yang}, {Yang}, {Yap}, {Yazback}, {Yeeles}, {Yu},
  {Yu}, {Yuen}, {Zadro{\.Z}ny}, {Zadro{\.Z}ny}, {Zanolin}, {Zelenova},
  {Zendri}, {Zevin}, {Zhang}, {Zhang}, {Zhang}, {Zhao}, {Zhao}, {Zhou}, {Zhou},
  {Zhu}, {Zimmerman}, {Zucker}, {Zweizig}, {LIGO Scientific Collaboration}, \&
  {Virgo Collaboration}}]{abbott_etal2020}
{Abbott}, R., {Abbott}, T.~D., {Abraham}, S., {et~al.} 2020, \prl, 125, 101102

\bibitem[{{Anderson} \& {King}(2003)}]{anderson_king2003}
{Anderson}, J. \& {King}, I.~R. 2003, \aj, 126, 772

\bibitem[{Arnold(1989)}]{arnold1989mathematical}
Arnold, V. 1989, Mathematical methods of classical mechanics, Vol.~60
  (Springer)

\bibitem[{{Ba{\~n}ados} {et~al.}(2018){Ba{\~n}ados}, {Venemans},
  {Mazzucchelli}, {Farina}, {Walter}, {Wang}, {Decarli}, {Stern}, {Fan},
  {Davies}, {Hennawi}, {Simcoe}, {Turner}, {Rix}, {Yang}, {Kelson}, {Rudie}, \&
  {Winters}}]{Banados_etal2018}
{Ba{\~n}ados}, E., {Venemans}, B.~P., {Mazzucchelli}, C., {et~al.} 2018, \nat,
  553, 473

\bibitem[{{Baumgardt} \& {Hilker}(2018)}]{Baumgardt_Hilker_2018}
{Baumgardt}, H. \& {Hilker}, M. 2018, \mnras, 478, 1520

\bibitem[{{Bellini} {et~al.}(2017){Bellini}, {Bianchini}, {Varri}, {Anderson},
  {Piotto}, {van der Marel}, {Vesperini}, \& {Watkins}}]{bellini_etal2017}
{Bellini}, A., {Bianchini}, P., {Varri}, A.~L., {et~al.} 2017, \apj, 844, 167

\bibitem[{{Binney} \& {Piffl}(2015)}]{BinneyPiffl2015}
{Binney}, J. \& {Piffl}, T. 2015, \mnras, 454, 3653

\bibitem[{{Binney} \& {Tremaine}(2008)}]{BT2008}
{Binney}, J. \& {Tremaine}, S. 2008, {Galactic Dynamics: Second Edition}
  (Princeton University Press, Princeton, NJ)

\bibitem[{{de Boer} {et~al.}(2019){de Boer}, {Gieles}, {Balbinot},
  {H{\'e}nault-Brunet}, {Sollima}, {Watkins}, \& {Claydon}}]{deBoer_etal2019}
{de Boer}, T.~J.~L., {Gieles}, M., {Balbinot}, E., {et~al.} 2019, \mnras, 485,
  4906

\bibitem[{{de Rijcke} {et~al.}(2006){de Rijcke}, {Buyle}, \&
  {Dejonghe}}]{deRijcke_etal2006}
{de Rijcke}, S., {Buyle}, P., \& {Dejonghe}, H. 2006, \mnras, 368, L43

\bibitem[{{den Brok} {et~al.}(2015){den Brok}, {Seth}, {Barth}, {Carson},
  {Neumayer}, {Cappellari}, {Debattista}, {Ho}, {Hood}, \&
  {McDermid}}]{Brok_2015}
{den Brok}, M., {Seth}, A.~C., {Barth}, A.~J., {et~al.} 2015, \apj, 809, 101

\bibitem[{{Dickson} {et~al.}(2023){Dickson}, {H{\'e}nault-Brunet}, {Baumgardt},
  {Gieles}, \& {Smith}}]{Dickson_etal2023}
{Dickson}, N., {H{\'e}nault-Brunet}, V., {Baumgardt}, H., {Gieles}, M., \&
  {Smith}, P.~J. 2023, \mnras, 522, 5320

\bibitem[{{Evans} \& {Williams}(2014)}]{evans2014}
{Evans}, N.~W. \& {Williams}, A.~A. 2014, \mnras, 443, 791

\bibitem[{{Foreman-Mackey} {et~al.}(2013){Foreman-Mackey}, {Hogg}, {Lang}, \&
  {Goodman}}]{emcee_package}
{Foreman-Mackey}, D., {Hogg}, D.~W., {Lang}, D., \& {Goodman}, J. 2013, \pasp,
  125, 306

\bibitem[{{Freire} {et~al.}(2017){Freire}, {Ridolfi}, {Kramer}, {Jordan},
  {Manchester}, {Torne}, {Sarkissian}, {Heinke}, {D'Amico}, {Camilo},
  {Lorimer}, \& {Lyne}}]{Freire_etal2017}
{Freire}, P.~C.~C., {Ridolfi}, A., {Kramer}, M., {et~al.} 2017, \mnras, 471,
  857

\bibitem[{{Gerssen} {et~al.}(2002){Gerssen}, {van der Marel}, {Gebhardt},
  {Guhathakurta}, {Peterson}, \& {Pryor}}]{Gerssen_etal2002}
{Gerssen}, J., {van der Marel}, R.~P., {Gebhardt}, K., {et~al.} 2002, \aj, 124,
  3270

\bibitem[{{Gieles} \& {Zocchi}(2015)}]{GielesZocchi_2015}
{Gieles}, M. \& {Zocchi}, A. 2015, \mnras, 454, 576

\bibitem[{{Giersz} {et~al.}(2015){Giersz}, {Leigh}, {Hypki}, {L{\"u}tzgendorf},
  \& {Askar}}]{Giersz_etal2015}
{Giersz}, M., {Leigh}, N., {Hypki}, A., {L{\"u}tzgendorf}, N., \& {Askar}, A.
  2015, \mnras, 454, 3150

\bibitem[{{Goldsbury} {et~al.}(2010){Goldsbury}, {Richer}, {Anderson},
  {Dotter}, {Sarajedini}, \& {Woodley}}]{Goldsbury+2010}
{Goldsbury}, R., {Richer}, H.~B., {Anderson}, J., {et~al.} 2010, \aj, 140, 1830

\bibitem[{{Greene} {et~al.}(2020){Greene}, {Strader}, \&
  {Ho}}]{Greene_etal2020}
{Greene}, J.~E., {Strader}, J., \& {Ho}, L.~C. 2020, \araa, 58, 257

\bibitem[{{H{\'e}nault-Brunet} {et~al.}(2020){H{\'e}nault-Brunet}, {Gieles},
  {Strader}, {Peuten}, {Balbinot}, \& {Douglas}}]{HenaultBrunet_etal2020}
{H{\'e}nault-Brunet}, V., {Gieles}, M., {Strader}, J., {et~al.} 2020, \mnras,
  491, 113

\bibitem[{{Kamann} {et~al.}(2018){Kamann}, {Husser}, {Dreizler}, {Emsellem},
  {Weilbacher}, {Martens}, {Bacon}, {den Brok}, {Giesers}, {Krajnovi{\'c}},
  {Roth}, {Wendt}, \& {Wisotzki}}]{kamann+18}
{Kamann}, S., {Husser}, T.~O., {Dreizler}, S., {et~al.} 2018, \mnras, 473, 5591

\bibitem[{{K{\i}z{\i}ltan} {et~al.}(2017){K{\i}z{\i}ltan}, {Baumgardt}, \&
  {Loeb}}]{Kiziltan_etal2017}
{K{\i}z{\i}ltan}, B., {Baumgardt}, H., \& {Loeb}, A. 2017, \nat, 542, 203

\bibitem[{{Libralato} {et~al.}(2022){Libralato}, {Bellini}, {Vesperini},
  {Piotto}, {Milone}, {van der Marel}, {Anderson}, {Aparicio}, {Barbuy},
  {Bedin}, {Borsato}, {Cassisi}, {Dalessandro}, {Ferraro}, {King}, {Lanzoni},
  {Nardiello}, {Ortolani}, {Sarajedini}, \& {Sohn}}]{libralato+22}
{Libralato}, M., {Bellini}, A., {Vesperini}, E., {et~al.} 2022, \apj, 934, 150

\bibitem[{{Magorrian} {et~al.}(1998){Magorrian}, {Tremaine}, {Richstone},
  {Bender}, {Bower}, {Dressler}, {Faber}, {Gebhardt}, {Green}, {Grillmair},
  {Kormendy}, \& {Lauer}}]{Magorrian_etal1998}
{Magorrian}, J., {Tremaine}, S., {Richstone}, D., {et~al.} 1998, \aj, 115, 2285

\bibitem[{{Mann} {et~al.}(2020){Mann}, {Richer}, {Heyl}, {Anderson}, {Kalirai},
  {Caiazzo}, {M{\"o}hle}, {Knee}, \& {Baumgardt}}]{Mann_etal2020}
{Mann}, C.~R., {Richer}, H., {Heyl}, J., {et~al.} 2020, \apj, 893, 86

\bibitem[{{McLaughlin} {et~al.}(2006){McLaughlin}, {Anderson}, {Meylan},
  {Gebhardt}, {Pryor}, {Minniti}, \& {Phinney}}]{mclaughlin_etal2006}
{McLaughlin}, D.~E., {Anderson}, J., {Meylan}, G., {et~al.} 2006, \apjs, 166,
  249

\bibitem[{{Miller} \& {Hamilton}(2002)}]{MillerHamilton_2002}
{Miller}, M.~C. \& {Hamilton}, D.~P. 2002, \mnras, 330, 232

\bibitem[{{Miocchi} {et~al.}(2013){Miocchi}, {Lanzoni}, {Ferraro},
  {Dalessandro}, {Vesperini}, {Pasquato}, {Beccari}, {Pallanca}, \&
  {Sanna}}]{Miocchi_etal2013}
{Miocchi}, P., {Lanzoni}, B., {Ferraro}, F.~R., {et~al.} 2013, \apj, 774, 151

\bibitem[{{Nelson} {et~al.}(2014){Nelson}, {Ford}, \&
  {Payne}}]{nelson_etal2014}
{Nelson}, B., {Ford}, E.~B., \& {Payne}, M.~J. 2014, \apjs, 210, 11

\bibitem[{{Nguyen} {et~al.}(2018){Nguyen}, {Seth}, {Neumayer}, {Kamann},
  {Voggel}, {Cappellari}, {Picotti}, {Nguyen}, {B{\"o}ker}, {Debattista},
  {Caldwell}, {McDermid}, {Bastian}, {Ahn}, \& {Pechetti}}]{Nguyen_etal2018}
{Nguyen}, D.~D., {Seth}, A.~C., {Neumayer}, N., {et~al.} 2018, \apj, 858, 118

\bibitem[{{Pascale} {et~al.}(2019){Pascale}, {Binney}, {Nipoti}, \&
  {Posti}}]{Pascale_etal2019}
{Pascale}, R., {Binney}, J., {Nipoti}, C., \& {Posti}, L. 2019, \mnras, 488,
  2423

\bibitem[{{Pascale} {et~al.}(2018){Pascale}, {Posti}, {Nipoti}, \&
  {Binney}}]{Pascale_etal2018}
{Pascale}, R., {Posti}, L., {Nipoti}, C., \& {Binney}, J. 2018, \mnras, 480,
  927

\bibitem[{{Piffl} {et~al.}(2015){Piffl}, {Penoyre}, \&
  {Binney}}]{Piffl_etal2015}
{Piffl}, T., {Penoyre}, Z., \& {Binney}, J. 2015, \mnras, 451, 639

\bibitem[{{Portegies Zwart} {et~al.}(2004){Portegies Zwart}, {Baumgardt},
  {Hut}, {Makino}, \& {McMillan}}]{Portegies-Zwart_etal2004}
{Portegies Zwart}, S.~F., {Baumgardt}, H., {Hut}, P., {Makino}, J., \&
  {McMillan}, S. L.~W. 2004, \nat, 428, 724

\bibitem[{{Posti} {et~al.}(2015){Posti}, {Binney}, {Nipoti}, \&
  {Ciotti}}]{Posti_etal2015}
{Posti}, L., {Binney}, J., {Nipoti}, C., \& {Ciotti}, L. 2015, \mnras, 447,
  3060

\bibitem[{{Ridolfi} {et~al.}(2016){Ridolfi}, {Freire}, {Torne}, {Heinke}, {van
  den Berg}, {Jordan}, {Kramer}, {Bassa}, {Sarkissian}, {D'Amico}, {Lorimer},
  {Camilo}, {Manchester}, \& {Lyne}}]{ridolfi_etal2016}
{Ridolfi}, A., {Freire}, P.~C.~C., {Torne}, P., {et~al.} 2016, \mnras, 462,
  2918

\bibitem[{{Strader} {et~al.}(2012){Strader}, {Chomiuk}, {Maccarone},
  {Miller-Jones}, {Seth}, {Heinke}, \& {Sivakoff}}]{Strader_etal2012}
{Strader}, J., {Chomiuk}, L., {Maccarone}, T.~J., {et~al.} 2012, \apjl, 750,
  L27

\bibitem[{{ter Braak} \& {Vrugt}(2008)}]{tarBraak_etal2008}
{ter Braak}, C. \& {Vrugt}, J. 2008, Statistics and Computing, 18, 435

\bibitem[{{Trager} {et~al.}(1995){Trager}, {King}, \&
  {Djorgovski}}]{Trager_etal1995}
{Trager}, S.~C., {King}, I.~R., \& {Djorgovski}, S. 1995, \aj, 109, 218

\bibitem[{{Tremou} {et~al.}(2018){Tremou}, {Strader}, {Chomiuk}, {Shishkovsky},
  {Maccarone}, {Miller-Jones}, {Tudor}, {Heinke}, {Sivakoff}, {Seth}, \&
  {Noyola}}]{Tremou_2018}
{Tremou}, E., {Strader}, J., {Chomiuk}, L., {et~al.} 2018, \apj, 862, 16

\bibitem[{{Vasiliev}(2019)}]{vasiliev_2019}
{Vasiliev}, E. 2019, \mnras, 482, 1525

\bibitem[{{Vitral} {et~al.}(2023){Vitral}, {Libralato}, {Kremer}, {Mamon},
  {Bellini}, {Bedin}, \& {Anderson}}]{vitral_etal2023}
{Vitral}, E., {Libralato}, M., {Kremer}, K., {et~al.} 2023, arXiv e-prints,
  arXiv:2305.12702

\bibitem[{{Volonteri}(2010)}]{Volonteri_2010}
{Volonteri}, M. 2010, \aapr, 18, 279

\end{thebibliography}

\appendix
\section{Model-data comparison}\label{sec:posterior_definition}
We inferred the model's free parameters in a Bayesian framework. In particular, defining the vector of 11 free parameters $\boldsymbol{\theta} \equiv \{ \log M_\star, \log J_0, \zeta, \Gamma, B, g_z, h_z, \log J_{\rm cut}, \alpha, \log M_\bullet, m \}$, namely the nine DF parameters (see eq.~\ref{eq:DF}), the logarithm of the IMBH mass ($\log M_\bullet$), and the normalization factor of the density profile ($m$),  we can write the posterior distribution as
\begin{equation}
    p(\boldsymbol{\theta} | \boldsymbol{D}) \propto p(\boldsymbol{\theta})\,p(\boldsymbol{D} | \boldsymbol{\theta})\,,
\end{equation}
where $\boldsymbol{D}$ is the data vector, including both the kinematic sample and the surface density profile.
The $p(\boldsymbol{\theta})$, and $p(\boldsymbol{D} | \boldsymbol{\theta}) \equiv \mathcal{L}(\boldsymbol{D})$ terms on the right-hand side of the equation are, respectively, the prior on the free parameters and the likelihood.

Assuming that all the data sets are independent of each other, we decompose the logarithm of the likelihood into the sum of the different terms
\begin{equation}
    \ln \mathcal{L}(\boldsymbol{D}) = 
    \ln \mathcal{L}_{v}+\ln \mathcal{L}_{\sigma_{\rm R}} + \ln \mathcal{L}_{\sigma_{\rm T}} + \ln \mathcal{L}_{\sigma_{\rm LOS}} +\ln \mathcal{L}_{n} \,.
    \label{eq:total_likelihood}
\end{equation}
We stress that for the cluster central region, we adopt a star-by-star approach modeling the velocity and error of each of the $N_{\rm stars}$ stars. The resulting likelihood is
\begin{equation}
    \ln \mathcal{L}_{v} \equiv \sum^{N_{\rm stars}}_{j=1} \ln \mathcal{F}(\boldsymbol{v}_j|R_j\,\delta\boldsymbol{v}_j)\,,
    \label{eq:likelihood_singlestars}
\end{equation}
where $\mathcal{F}\equiv \mathcal{V} \ast \mathcal{N}$. 
Therefore, for each star $j$, the velocity distribution ($\mathcal{V}$, see eq.~\ref{eq:3Dprojected_veldistr} to \ref{eq:PMVD}, and computed at the observed projected distance $R_j$) is convolved ($\ast$) with observational errors, represented by a zero-mean, multivariate Gaussian ($\mathcal{N}$).
The Gaussian covariance matrix has diagonal elements equal to the errors squared, $\delta\boldsymbol{v}^2_j$, and zero off-diagonal terms.
% with dispersion $\delta\boldsymbol{v}_j$.
The resulting function is evaluated at the observed velocity ($\boldsymbol{v}_j$). 
When the full kinematic information (PM and LOS velocity) is not available for the $j$-th star, we marginalize over the missing velocity components (see eq.~\ref{eq:LOSVD}, and \ref{eq:PMVD}).

For the velocity dispersions outside the central 12", we defined
\begin{equation}
    \ln \mathcal{L}_{\sigma_{i}} \equiv  -\frac{1}{2} \sum^{N_{\rm bin,}i}_{k=1}  \frac{(\sigma_{i,\,k} - \sigma_{\star\,i}(R_{k}) )^2}{\delta\sigma_{i,\,k}^2} \quad {i}\in\{\rm R, T, LOS\} \,,
    \label{eq:likelihood_veldisp}
\end{equation}
with $\sigma_{i,\,k}$ 
being the velocity dispersion of the $i$-th component in the $k$-th radial bin (centered in $R_{\rm k}$), $\delta\sigma_i{,\,k}$ the corresponding error, and $\sigma_{\star\,i}(R_{k})$ the model prediction. $N_{\rm bin,}i$ is the total number of bins in which the velocity dispersion was obtained.
Similarly, for the density profile
\begin{equation}
    \ln \mathcal{L}_{n} \equiv  -\frac{1}{2} \sum^{N_{\rm prof}}_{l=1} \frac{(n_l - n_{\star}(R_l) )^2}{\delta n_l^2}\,,
    \label{eq:likelihood_densityprofile}
\end{equation}
where $N_{\rm prof}$ is the total number of bins in the surface density profile.

\begin{figure*}
    \centering
    \includegraphics[width=\textwidth]{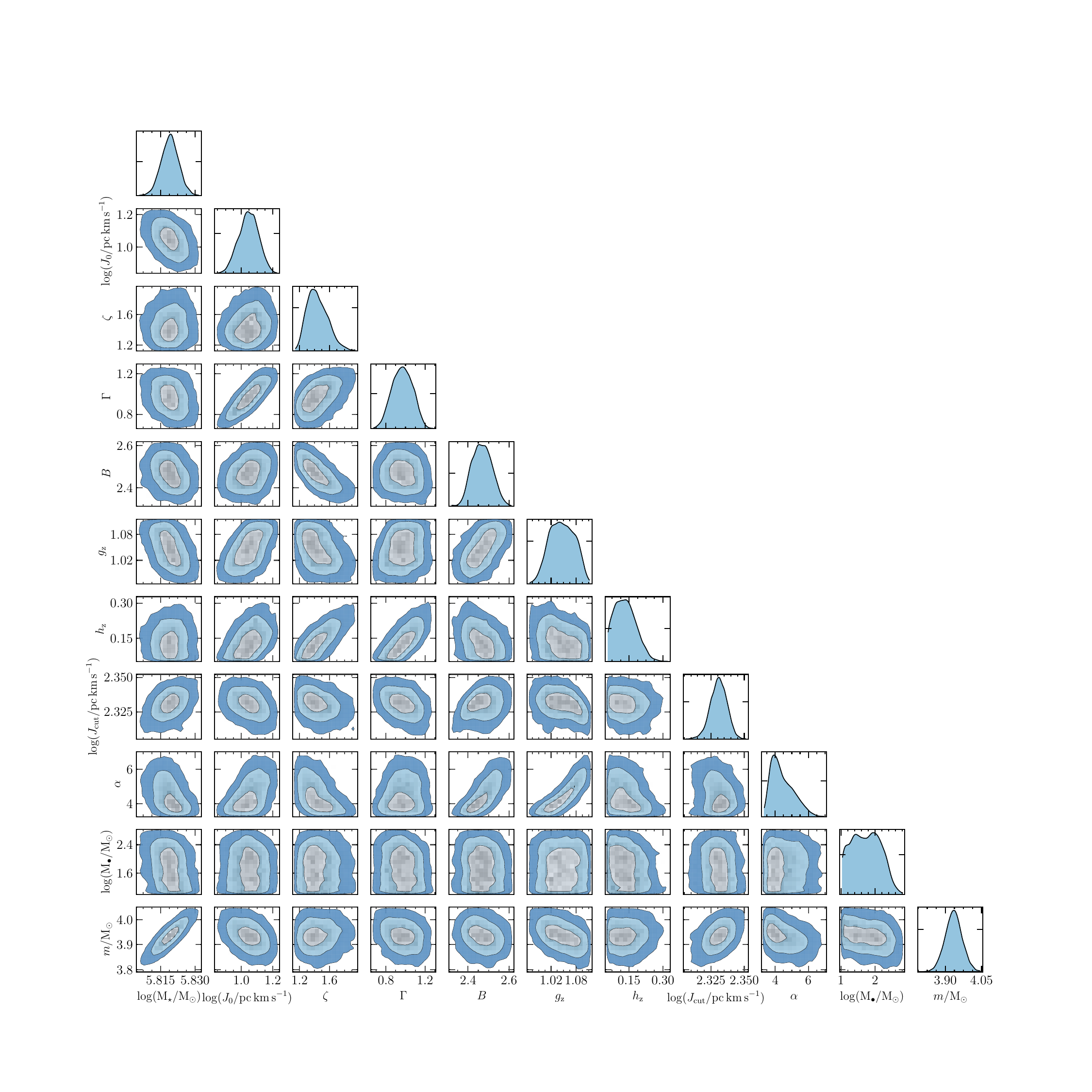}
    \caption{ 
    One- (diagonal panels) and two-dimensional (lower-diagonal panels), marginalized, posterior distributions over the model's free parameters.
    % The vertical lines in each diagonal panel show the 16th, 50th, and 84th percentiles.
    % Median values as well as 16th and 84th percentiles (quoted as errors) are shown at the top of each diagonal panel.
    See Section~\ref{sec:models} for a description of the parameters. 
    Prior ranges, median values, 68\%, and 99.7\% CIs for each parameter are reported in Table~\ref{tab:prior_posterior_values}.
    }
\label{fig:corner_plot}
\end{figure*}

All the physical quantities of the model are self-consistently computed from the DF (see Section~\ref{sec:observables_from_DF}).
For all the parameters we assumed uniform priors (Table~\ref{tab:prior_posterior_values} for the specific prior ranges adopted).
In particular, for the IMBH mass, we adopted a lower limit of 10 M$_\odot$, well below the nominal definition of IMBH. Also, a less massive BH would have $R_{\rm infl}<10^{-3}$ pc with negligible impact on observables.
We explored the free-parameter space by means of a Markov Chain Monte Carlo algorithm (MCMC), using the \texttt{emcee} Python package \citep{emcee_package}. 
The algorithm was run with 112 walkers for about 7,000 steps each. 
We used a mixture of the \emph{moves} developed by
\citet{tarBraak_etal2008}, and \citet{nelson_etal2014}, to achieve a more efficient exploration of the parameters' space.
For each walker, we discarded the first 2500 steps to account for the initial convergence phase, while exploring the prior range.
After that, we accounted for the correlation between subsequent samples taking one sample every 100.
Finally, we obtained about 5000 independent posterior samples.
In Fig.~\ref{fig:corner_plot}, we show the corner plot with both the marginalized posterior distributions (diagonal panels) and two-dimensional joint distributions (lower-diagonal panels).

In Table~\ref{tab:prior_posterior_values} we list the parameter prior ranges and the posterior values from the MCMC fitting.
For each model free parameter, we report the median value, and the 68\% ($1\sigma$) and 99.7\% ($3\sigma$) CIs, computed from posterior samples.
We note that all the free parameters are well constrained within the prior ranges. This indicates that the adopted intervals were well suited for a thorough exploration of the free-parameters space, and there is no evidence for the need to enlarge these ranges.
\renewcommand{\arraystretch}{1.95} % Default value: 1

\begin{table*}[]
    \centering
    \begin{tabular}{lcccc}
    \hline
    Parameter & Prior range & \SetCell[c=3]{} & Posterior & \\
    \cline{3-5}
    & & Median & 68\% CI & 99.7\% CI \\
    \hline \hline
    $\log$(M$_\star$/M$_\odot$) & [5.0; 7.0]                & 5.819 & [5.815; 5.823] & [5.807; 5.831]\\
    $\log (J_0/{\rm pc\,km\,s}^{-1})$ & [0.3; 1.5]          & 1.05 & [0.98; 1.11] & [0.87; 1.2]\\
    $\zeta$ & [0.5; 5.0]                                    & 1.43 & [1.31; 1.59] & [1.15; 1.96]\\
    $\Gamma$ & [0.0; 2.0]                                   & 0.97 & [0.86; 1.08] & [0.69; 1.22]\\
    $B$ & [1.0; 4.0]                                        & 2.46 & [2.42; 2.51] & [2.32; 2.6]\\
    $g_{\rm z}$ & [0.05; 1.45]                              & 1.04 & [1.01; 1.08] & [0.97; 1.11]\\
    $h_{\rm z}$ & [0.05; 1.45]                              & 0.13 & [0.08; 0.18] & [0.05; 0.29]\\
    $\log (J_{\rm cut}/{\rm pc\,km\,s}^{-1})$ & [1.5; 3.0]  & 2.331 & [2.325; 2.337] & [2.311; 2.348]\\
    $\alpha$ & [2.0; 10.0]                                  & 4.23 & [3.75; 5.04] & [3.27; 6.55]\\
    $\log$(M$_\bullet$/M$_\odot$) & [1.0; 5.5]              & 1.74 & [1.27; 2.19] & [1.0; 2.76]\\
    $m/$M$_\odot$ & [2.0; 7.0]                              & 3.93 & [3.9; 3.97] & [3.83; 4.04]\\
    \hline
    \end{tabular}
    \caption{Free parameters of the model as defined in Section~\ref{sec:models}.
    The central column shows the adopted prior ranges in each parameter.
    The rightmost columns report the median values,  the 68\%, and 99.7\% CIs of the posterior distributions.
    We note that the $3\sigma$ lower limit on the IMBH mass coincides with the lower boundary of the prior (10~M$_\odot$). 
    }
    \label{tab:prior_posterior_values}
\end{table*}

\end{document}